\documentclass[10pt]{elsarticle}

\usepackage{hyperref,color,amssymb}
\makeatletter
\def\ps@pprintTitle{%
	\let\@oddhead\@empty
	\let\@evenhead\@empty
	\let\@oddfoot\@empty
	\let\@evenfoot\@oddfoot
}
\makeatother
\begin{document}

\renewcommand{\baselinestretch}{1.45}
	
	\begin{frontmatter}
		
		\title{Performance limit for base-excited energy harvesting, and comparison with experiments}
		
		\author{Sankalp Tiwari\fnref{myfootnote1}}
		\address{Mechanical Engineering\\ Indian Institute of Technology Kanpur\\ Kanpur, 208016, India}
		\fntext[myfootnote1]{Email: snklptwr@gmail.com, sankalpt@iitk.ac.in}
		
		\author{C. P. Vyasarayani\fnref{myfootnote2}\corref{mycorrespondingauthor}}
		\address{Mechanical and Aerospace Engineering\\ Indian Institute of Technology Hyderabad \\
			Sangareddy, 502285, India}
		\fntext[myfootnote2]{Email: vcprakash@mae.iith.ac.in}
		\cortext[mycorrespondingauthor]{Corresponding author}
		
		\author{Anindya Chatterjee\fnref{myfootnote3}}
		\address{Mechanical Engineering\\ Indian Institute of Technology Kanpur\\ Kanpur, 208016, India}
		\fntext[myfootnote3]{Email: anindya100@gmail.com, anindya@iitk.ac.in}
		
		\begin{abstract}
			We consider the theoretical maximum extractable average power from an energy harvesting device attached to a vibrating table which provides a unidirectional displacement $A\sin(\omega t)$. The total mass of moving components in the device is $m$. The device is assembled in a container of dimension $L$, limiting the displacements and deformations of components within. The masses in the device may be interconnected in arbitrary ways. The maximum extractable average power is bounded by $\frac{mLA\omega^3}{\pi}$, for motions in 1, 2, or 3 dimensions; with both rectilinear and rotary motions as special cases; and with either single or multiple degrees of freedom. The limiting displacement profile of the moving masses for extracting maximum power is discontinuous, and not physically realizable. But smooth approximations can be nearly as good: with $15$ terms in a Fourier approximation, the upper limit is $99$\% of the theoretical maximum. Purely sinusoidal solutions are limited to $\frac{\pi}{4}$ times the theoretical maximum. For both single-degree-of-freedom linear resonant devices and nonresonant  devices where the energy extraction mimics a linear torsional damper, the maximum average power output is $\frac{mLA\omega^3}{4}$. Thirty-six experimental energy harvesting devices in the literature are found to extract power amounts ranging from $0.0036$\% to $29$\% of the theoretical maximum. Of these thirty-six, twenty achieve less than 2\% and three achieve more than 20\%. We suggest, as a figure of merit, that energy extraction above $\frac{0.2 mLA\omega^3}{\pi}$ may be considered excellent, and extraction above $\frac{0.3 mLA\omega^3}{\pi}$ may be considered challenging.
		\end{abstract}
		
		
	\end{frontmatter}
	
	
	\section{Introduction}
	\label{intro}

Dynamic energy harvesting has been an important research area in recent decades. There is an abundant literature on such energy harvesting devices. Several important review papers have appeared in the last decade.

Kim et al~\cite{kim2011review}, in 2011, reviewed 93 papers on piezoelectric devices that harvest energy from vibration. They specifically mention the need for the development of flexible and resilient piezoelectric materials with a high coupling coefficient and better fatigue life. They conclude that ``new efficient circuitry for energy harvesters is necessary" to improve efficiency.

In 2013, Harne and Wang~\cite{harne2013review} reviewed 84 papers that used bi-stability as a means to achieve large vibrations. They mention that in most devices, bi-stability is induced by magnetic forces or mechanical loading. They concluded that nonlinear bi-stable energy harvesters produce an order of magnitude more power than their linear counterparts, at least in steady-state operation.

Also in 2013, Pelegrini et al~\cite{pellegrini2013bistable} reviewed 50 papers on bi-stable energy harvesters. They classified the devices based on whether the bi-stability is used directly for energy harvesting or for frequency up-conversion.  They noted that a Duffing oscillator like model is widely used to study bi-stable harvesters. Out of intra-well, chaotic, and large amplitude limit cycle responses, the latter is preferred as it generates more power and simplifies the needed electric circuitry. They finally proposed a dimensionless metric
$$I=P_\textrm{rms}f/m a_{\textrm{rms}}^2$$
to compare different energy harvesters. Here $P_{\textrm{rms}}$ is the RMS power generated; $a_{\textrm{rms}}$ is the RMS base acceleration; $f$ is the frequency of excitation; and $m$ is the effective mass. If the device operates over a specific frequency interval $[a,\,b]$, they suggested using the index $$I_{a-b}=\frac{1}{b-a} \intop_{a}^{b}Idf.$$
One shortcoming of the above two metrics is that they do not limit the motion amplitudes of the vibrating devices in the harvester. In principle, at resonance, extremely lightly damped systems with very large amplitudes of vibration could extract very large amounts of power (we will discuss the lightly damped linear resonator later in this paper).

In 2014, Daqaq et al~\cite{daqaq2014role} reviewed 119 papers that exploit nonlinearities in the energy harvesters for performance enhancement.  The most critical future direction suggested by the authors was the need to define a performance metric for nonlinear energy harvesters. As almost direct motivation for the present paper, they wrote:
\begin{quote}
Nonuniqueness \ldots  aperiodicity and
	bifurcations \ldots as parameters vary \ldots  make developing direct performance metrics
	\ldots a challenging task. Therefore, developing
	such metrics represents an essential first step towards understanding
	the role of nonlinearity in the transduction of energy
	harvesters.
\end{quote}

In 2018, Tran et al~\cite{tran2018ambient} reviewed 183 papers on harvester designs that use nonlinear techniques for performance improvement. They compared the advantages and disadvantages of using different nonlinear designs in the harvesters. Designs examined had mono-, bi-, tri- and quad-stability; had stoppers, internal resonances, and parametric excitation; and used multimodal arrays, multi-degree of freedom oscillators, and coupled-mode techniques for bandwidth enhancement. Stochastic and pre-loading methods were also discussed for extracting energy at low excitation levels.

From the above review papers, it is clear that the topic of vibratory energy harvesting is technologically important and has been studied by many authors. However, it also appears that elementary upper bounds are not available on achievable power harvesting levels for arbitrary deformable multi-degree-of-freedom systems. Moreover, in our work we have realized  that the same upper bounds apply on extractable power, whether the internal motions are predominantly vibratory or include rotary or whirling components. We now review some papers that address rotary or whirling devices in energy harvesting.

\subsection{Rotary devices under translatory base excitation}
The commonest treatments of pendulums used for energy extraction under translatory base excitation are for energy from waves, e.g., see
Yurchenko and Alevras~\cite{yurchenko2018parametric} and
Sequeira et al~\cite{sequeira2019investigating}. However, such rotary devices can obviously be used at much smaller scales (centimeters or even millimeters instead of meters), at higher frequencies, and for energy sources other than ocean waves.
For example,
Lie~\cite{lei2010kinematic} et al. performed a kinematic analysis of an auto-winding system with a pawl-lever mechanism for converting oscillations into continuous rotation. This mechanism is similar to one used in some auto-winding wristwatches.

Specific performance aspects of pendulum-based energy extraction devices have been addressed by many authors, and a few representative examples are now given.
Lee and Chung~\cite{lee2016design} studied two pendulums mounted on the same shaft, parametrically excited in the horizontal direction. One pendulum carried a coil while the other carried a magnet. The best performance was obtained during anti-phase motion of the pendulums.
In Dotti et al~\cite{dotti2017damping}, the importance of accurately modeling the damping in parametric rotary pendulum was emphasized. Linear, quadratic, and Coulumb's dry friction terms were included to obtain an accurate match between modeling and experiments.
Marszal et al~\cite{marszal2017energy} proposed a parametric pendulum with a ratchet mechanism that uses pendulum oscillations to drive a generator. The best performance was achieved in 2:1 parametric resonance.
Simeone et al~\cite{simeone2019level} considered a base excited, horizontal, oscillating pendulum attached to a gearbox. They showed that higher performance is possible if the load is varied with frequency.
Kecik and Mitura~\cite{kecik2020energy} showed that a pendulum-based energy harvester can be designed to simultaneously extract energy and quench vibrations of the main structure.

Some researchers have used active control strategies to enhance harvesting. These papers are less directly related to our work, because we are interested in an upper bound for what is achievable. However, for completeness, some representative examples are given. Reguera et al~\cite{reguera2016rotation} proposed a control strategy in which the length of a vertically excited pendulum is varied to achieve continuous rotations. Das and Wahi~\cite{das2016initiation} used time delay control to initiate and sustain full rotations in a parametric pendulum under vertical base excitation. Similarly, Firoozy et al~\cite{firoozy2019using} has used time delay control to improve the performance of a maglev energy harvester under base excitation.  

In a work more directly related to ours, for a base excited pendulum, Nandakumar et al~\cite{nandakumar2012optimum} found the optimal rotation profile for extracting maximum power using nonlinear optimization. They observed that when the rotations approach a discontinuous staircase-like profile, power extraction is greatest. Their single-degree-of-freedom results will be automatically included within our more general results below.

Motivated by the above literature review including the comments quoted from Dadaq et al~\cite{daqaq2014role}, in this paper, we will obtain a fundamental bound on the maximum possible power extracted  for a general class of multibody harvesters. We will then compare this theoretical maximum with reported energy extraction rates in thirty experimental studies in the literature. 

\subsection{Problem statement and main findings}	
Our energy harvesting device is allowed to have some sort of casing or container, whose properties (including mass) do not affect our analysis. We assume that deformations of the casing, if any, have no influence on energy harvesting. Inside the casing, there are moving and/or deforming components of total mass $m$. The size of the device limits the displacement range of these moving components to $L$. The base excitation as a function of time $t$ is taken to be $A \sin (\omega t)$, usually with $A \ll L$.	We are interested in the following question: given $m$, $L$, $A$ and $\omega$, what is the maximum theoretically possible average rate of work  done by the vibrating table on the device? This same average work rate provides an upper bound on the power that can possibly be harvested from the device.

Such a theoretical maximum is useful for assessing the efficiency of any given physical device. How much of this maximum power can actually be harvested depends on the design of the mechanical and electrical transduction components of the energy harvester, as well as on technological parameters such as size, material choice, and operating frequency. The theoretical maximum is known for single-mass devices, in both rectilinear oscillations and rotary whirling. Here we extend the result to multi-degree-of-freedom systems undergoing arbitrary motions, and then compare with published results on several experimentally realized devices.

For single-degree-of-freedom (SDOF) resonant devices in rectilinear sinusoidal motion, Williams and Yates~\cite{williams1996analysis} found the maximum power to be $\displaystyle \frac{mLA\omega^3}{4}$. For general nonlinear SDOF devices, Ramlan et al ~\cite{ramlan2010potential} found the maximum power to be $\displaystyle\frac{mLA\omega^3}{\pi}$. 
Here we will explain why the same upper limit of $\displaystyle \frac{mLA\omega^3}{\pi}$ applies to all devices with a total moving mass $m$ and motion range $L$, whether with one or more degrees of freedom; whether linear or nonlinear in dynamics; regardless of how the moving masses are interconnected; whether the motions are rectilinear, circular, more arbitrary and planar, or even spatial. Furthermore, we will examine thirty-six different energy harvesting devices reported in the literature and find that the power extracted, as a proportion of the theoretical maximum, ranges from 0.0036\% to 29\% of this limit. Of those thirty-six devices, twenty extract under
2\% of their theoretical maximum, and three extract more than 20\% of theirs.

Twenty-two of these thirty-six devices have also been studied in a recent and interesting paper by Blad and Tolou~\cite{blad2019efficiency}. Their approach is more technological and design related while ours is more fundamental. In particular, their proposed efficiency measure uses
$$W_{\textrm{max}}=\frac{1}{16}\rho_m VL_z A\omega^3,$$
where $\rho_m$ is the density of the moving mass $m$, $V$ is the total volume of the device, and $L_z$ (with their choice of axes) is the size of the device in the direction of base excitation. Clearly, their adopted parameters are relevant to the construction and deployment of the device. At the same time, if the same device with the same mass $m$ is placed in a container with twice the size, then in our view the energy extraction efficiency does not change, while their proposed efficiency measure decreases greatly. Nevertheless, we encourage the reader to see~\cite{blad2019efficiency}. That paper has useful technological discussions of design types including both bandwidth and
loss of efficiency associated with frequency up-conversion,  and interesting graphical groupings of different designs in two-dimensional plots.

	\section{Dimensional analysis and an elementary example}
	\label{dimen}
	The average power or rate of work ($W$) extracted from the harvester depends not only on the details of the design, but also on $m$, $L$, $A$, and $\omega$. We may write
	$$W = mL^2 \omega^3 f(m, L, A, \omega, \mbox{harvester design parameters}),$$
	where $mL^2 \omega^3$ has units of power and the function $f$ is necessarily dimensionless. If we consider a specific design, such as an oscillatory system, then the design parameters may, e.g., be a spring constant ($k>0$), a damping coefficient ($c \ge 0$), and a parameter characterizing the energy converter, which we denote by the abstract quantity $\mu$. Then
	\begin{equation}
	\label{Wdef}
	W = mL^2 \omega^3 f(m, L, A, \omega, k, c, \mu),
	\end{equation}
	and we have the following optimization problem:\\

	\noindent {\bf Problem 1:} Given $m$, $L$, $A$, and $\omega$, find the energy harvester parameters $k$, $c$, and $\mu$ so as to maximize $W$ in Eq.~(\ref{Wdef}).\\

	The maximizing $k$, $c$ and $\mu$ depend on the given quantities $m$, $L$, $A$, and $\omega$, and so when we insert those optimal values in Eq.~(\ref{Wdef}) we obtain
	$$W_{\rm max} = mL^2 \omega^3 f(m, L, A, \omega, k(m, L, A, \omega), c(m, L, A, \omega), \mu(m, L, A, \omega)) = mL^2 \omega^3 f_0(m, L, A, \omega),$$
	where $f_0$ is some other dimensionless function of just the original four parameters. That in turn means
	\begin{equation}
	\label{Wdef1}
	W_{\rm max} = mL^2 \omega^3 f_{1,\,\rm max} \left ( \frac{A}{L} \right ),
	\end{equation}
	because $m$, $L$, $A$, and $\omega$ can together form only one dimensionless group, $A/L$. The nature of the function $f_{1,\,\rm max}$ in Eq.~(\ref{Wdef1}) depends on the type of energy harvester under consideration. For example, the dependence may differ based on whether any springs present are linear or nonlinear, or whether the motion has a rotary (whirling) component, and on what the precise physics of the energy conversion device is. 
	
	A simple example here may help to indicate the wide range of possibilities. As an obviously suboptimal design example, consider a rectilinear periodic oscillatory design in which there is simply an energy conversion device that applies a resistive force on the mass, equivalent to a dashpot with coefficient $c$. 
	The mass is otherwise unsupported (or perhaps it also has a spring of vanishingly small stiffness to keep it centered on average, but too weak to require explicit modeling). We may call this suboptimal design a ``damper-only'' design. The equation of motion of the mass is (letting $y$ denote the displacement relative to the casing)
	$$m \ddot y + c \dot y = m \omega^2 A \sin(\omega t).$$
	We emphasize that the $c \dot y$ term above represents the effect of a hypothetical energy conversion device.
	
	The steady-state, zero-mean solution is
	\begin{equation}
	\label{xsol}
	y = - \frac{A \omega m}{c^2 + \omega^2 m^2} \left ( \omega m \sin(\omega t) + c \cos (\omega t \right) ). 
	\end{equation}
	The average energy dissipated per unit time gives an upper bound on  power extraction; and this bound equals the average of $c \dot y^2$, which is
	$$W_{\rm  damper \, only} = \frac{m A^2 \omega^3}{2} \frac{c \omega m}{c^2 + \omega^2 m^2}.$$
	Maximizing the above with respect to $c$ (which occurs for $c = \omega m$), we find
	\begin{equation}
	\label{damp}
	W_{\rm max,\,  damper \, only} =  \frac{m A^2 \omega^3}{4}.
	\end{equation}
	It remains to check that the displacement amplitude of the mass is less than $L/2$. To that end, Eq.~(\ref{xsol}) yields the amplitude $A/\sqrt{2}$ for the optimal $c$, which is less than $L/2$ by the assumption that $A \ll L$.
	Comparing Eq.~(\ref{damp}) with Eq.~(\ref{Wdef1}), we note that the function $f_{\rm 1,\,max}$ for this case is given by
	\begin{equation}
	\label{quad}
	f_{\rm 1,\,max, \, damper \, only} \left ( \frac{A}{L} \right ) = \frac{A^2}{4 L^2},
	\end{equation}
	which is tiny if $A \ll L$. With resonant designs, examples of which abound in the literature, one can do a lot better than Eq.~(\ref{quad}).\\
	
	\noindent {\bf Nondimensionalization strategy:} In the rest of this paper we will use the above nondimensionalization, implicit in Eq.~(\ref{Wdef1}), and take the total moving mass $m=1$, the motion limit $L=1$, the forcing frequency $\omega=1$, and the amplitude $A$ to be a free parameter, all in any consistent system of units we like. The maximum power output obtained will then (by Eq.~(\ref{Wdef1})) be merely some function of the free parameter $A$, i.e., $f_{\rm 1,\,max}(A)$, which we will consider to be effectively nondimensional. Later, to incorporate the {\em dimensional} quantities, we will replace $A$ by $\frac{A}{L}$ within $f_{\rm 1,\,max}$ and multiply by $mL^2 \omega^3$.

	The above strategy will simplify our presentation with no loss of generality. We now proceed to obtain the theoretical upper bound on extracted energy.
	
	\section{Energy harvesting bounds}
	\label{Rectilinear}
	The restricted case for a single point mass in rectilinear motion has been studied before~\cite{hosseinloo2015fundamental}. We will study one or more point masses, each undergoing arbitrary motion.
	
Since we wish to present a general upper bound, we need to be precise about what kind of system we are considering, and what kinds we are leaving out. To this end, we will first define the system qualitatively, then use the language of free body diagrams to make things precise, and finally extend our arguments to a broader class of systems.

\begin{figure}[h!]
		\centering{{\includegraphics[width=0.9\textwidth]{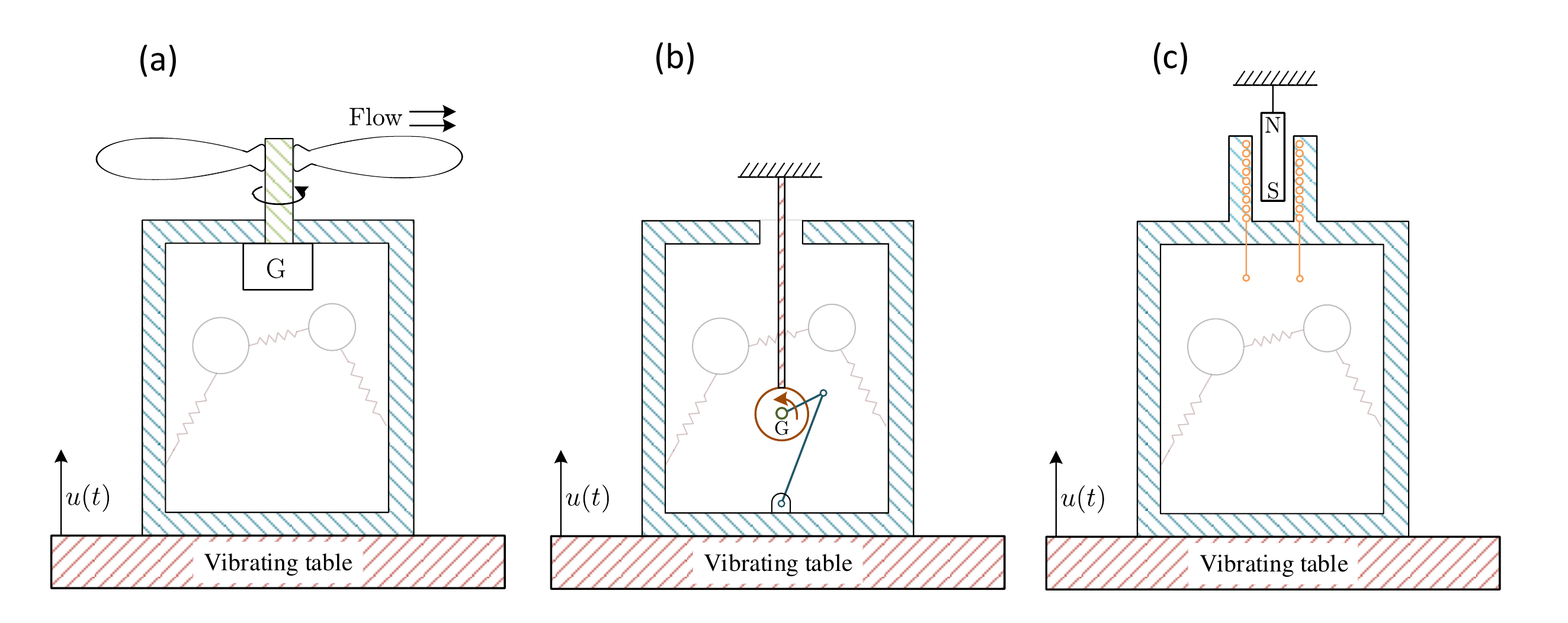}}}
		\caption{Examples of systems not included within our analysis. (a) External wind drives a fan, and nonzero energy is extracted with zero
base motion. (b) Access to an inertially-fixed external point allows a slider crank mechanism to extract unbounded energy.
(c) A permanent magnet attached to an inertially fixed point allows us to use an in-principle
massless coil to generate electricity.}
		\label{out}
	\end{figure}

To begin, note that we are talking about usual designs in energy harvesting, i.e., we consider a device that is mounted on the base, interacts with no other independently moving or stationary body or external material in ways that help generate energy, and extracts energy {\em solely} due to the motion of the vibrating base with respect to an inertial frame of reference. Moreover, the base motion is assumed to be
unaffected by the forces arising due to the energy harvesting device. The consequences that we wish to incorporate from these restrictions will be clear when we consider a free body diagram. However, it is helpful to consider first some examples of devices that are {\em not} included: see Fig.\ (\ref{out}) which shows a windmill, a slider-crank that connects to an external fixed point, and a massless harvester that extracts energy using external fields.

	\label{Nonlinaer}

In contrast to the systems in Fig.\ (\ref{out}), in the devices we include there are movable masses which move relative to the casing due to inertial effects induced by base motions. There are also electrical, magnetic, piezoelectric, or other energy-converting devices located completely within the casing, such that all action-reaction force pairs are internal to the casing.

\begin{figure}[h!]
		\centering{{\includegraphics[width=0.9\textwidth]{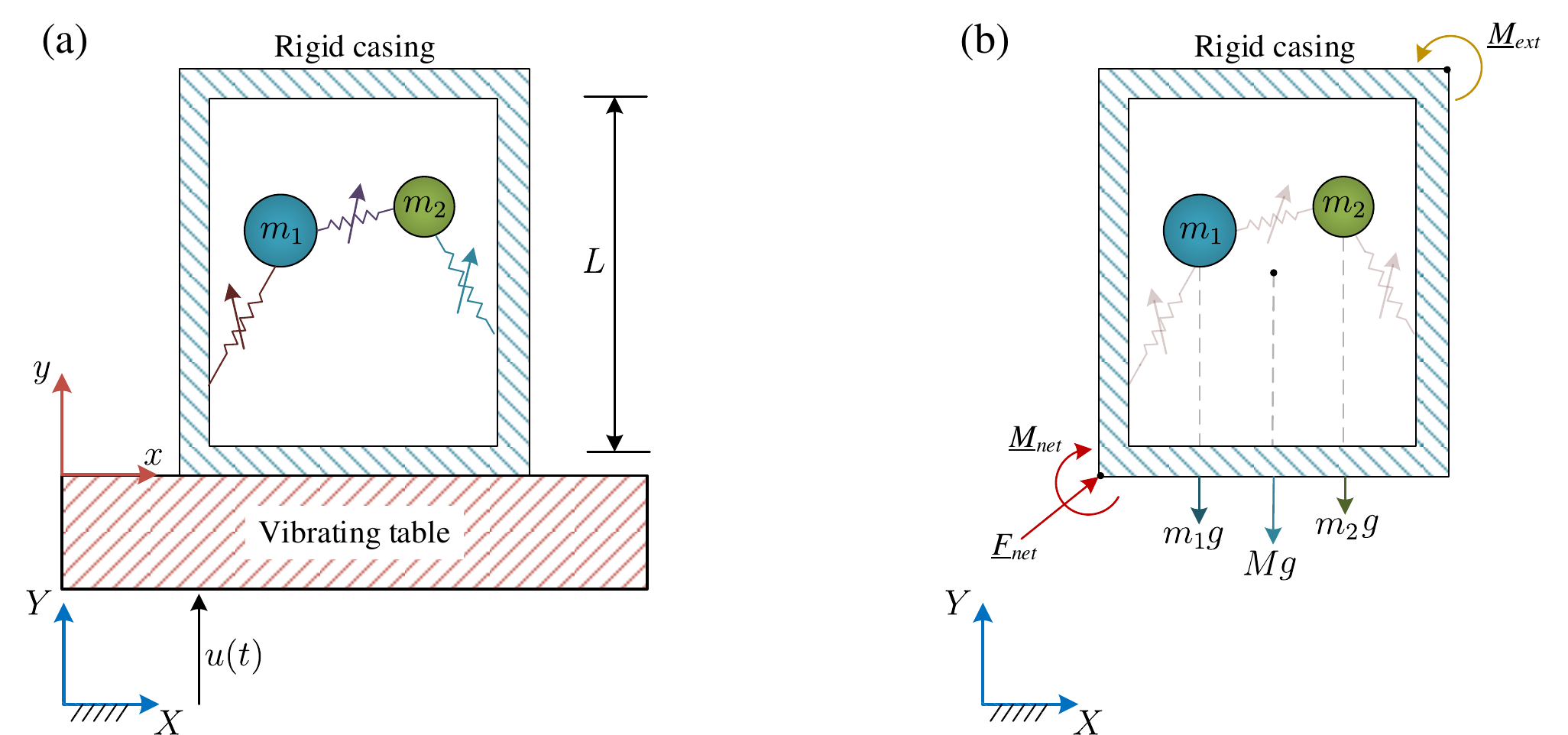}}}
		\caption{Schematic of a generic nonlinear energy harvester. The springs with arrows across them represent {\em any} massless connecting system whose forces follow the law of action and reaction, such that the {\em net} force drops out of the
free body diagram. (a) The device is attached to a vibrating base or table. (b) The free body diagram of the device shows forces and moments on the casing, and weights of the masses and the casing {\em only}. A reader may wonder why weights are included, because they
represent an interaction with an external inertial body (Earth); but it will be seen that weight does not lead to extraction of energy,
so there is no contradiction. Inclusion of weight makes our analysis clearly applicable to some large pendulum devices.}
		\label{Schematic_non}
	\end{figure}

Under the above assumptions, the average power extracted by the harvesting device cannot be greater than the average rate of mechanical work done by the base on the casing. With this view, we will consider the rate of work done by the moving base on the casing. Additionally, knowing that various unspecified forces related to energy conversion are now {\em internal forces}, we do not need to include them in a free body diagram:
see Fig.\ (\ref{Schematic_non}).

We now consider a base-excited mechanical  device enclosed in a container, which has length $L$ and mass $M$, as shown in Fig.~\ref{Schematic_non}(a). Two point masses $m_1$ and $m_2$ are internally connected to the device, and/or connected to each other, using arbitrary massless nonlinear elements (like springs, dashpots, linkages, etc.). It will be clear soon that these two point masses can in principle be replaced by an arbitrary number of masses. By the nondimensionalization scheme outlined earlier, we will take
$$m_1 + m_2 = 1.$$

Figure~\ref{Schematic_non}(b) shows the free body diagram of the device. The free body diagram includes a net force $\underline{F}_{\textrm{net}}$ and a net moment $\underline{M}_{\textrm{net}}$ from the table (see Fig.~\ref{Schematic_non}(a)). We also include a net external force
$$\underline{Q}=-(m_1g+m_2g+Mg)\,\textrm{j}$$
due to gravity, along with a possible external moment $\underline{M}_{\textrm{ext}}$ that can act from external sources on the casing. If there is any other force on the casing from the external environment, that force is treated as acting directly on the base and is not explicitly considered: this is because work done by the base in acting against that force occurs through loads transmitted {\em via} the casing, does not lead to internal extraction of energy, and does not affect our energy bound\footnote{%
	An example may help. Suppose we have a long vertical vibrating pipe in the ocean, and on this pipe we attach an energy
	harvesting device for some sensing application. The harvesting device changes the local hydrodynamics and may slightly
	change the local fluid forces
	experienced by the pipe, but those forces are external and not included in our study of the efficiency of the
	energy harvester.}.
We emphasize that, by the rules of drawing free body diagrams,
internal forces due to massless components and energy conversion devices are {\em not shown}.

The work done by the moving base on the casing depends on two things only: (i) the net force and moment from the base on the casing, and (ii) the motion of the casing. Since the base has no rotation, we must account for $\underline{F}_{\textrm{net}}$ alone in computing the work done. We can find $\underline{F}_{\textrm{net}}$ using linear momentum balance applied to the free body diagram,
	\begin{equation}
	\underline{F}_{\textrm{net}} = - \underline{Q} + \left[m_{1}\ddot{x}_{1}+m_{2}\ddot{x}_{2}\right]\hat{\textrm{i}}+\left[m_{1}(\ddot{y}_{1}+\ddot{u})+m_{2}(\ddot{y}_{2}+\ddot{u})+M\ddot{u}\right]\hat{\textrm{j}} + \left[m_{1}\ddot{z}_{1}+m_{2}\ddot{z}_{2}\right]\hat{\textrm{k}}.
	\label{Fnet}
	\end{equation}
	The average rate of work done by the vibrating base on the energy harvesting device due to $\underline{F}_{\textrm{net}}$, computed over some suitable
time interval $T$, is
	\begin{equation}
	W_{\rm two\, masses}=\frac{1}{T}\intop_{0}^{T}\underline{F}_{\textrm{net}}. \dot{u} \hat{\textrm{j}} \,dt.
	\label{NetWork}
	\end{equation}
The above can be found using Eq.\ (\ref{Fnet}), where the $x$- and $z$-direction motions drop out of the dot product. Since the base motion $u(t)$ is periodic, the contribution from the constant $\underline{Q}$ has zero average (the integral of $\dot u$ over one cycle is zero). Finally, note that the integral of $\ddot u \dot u$ over one cycle of $u$ is zero as well.
Therefore, we are left with
	\begin{equation}
	W_{\rm two\, masses}=\frac{1}{T}\intop_{0}^{T}\left[m_{1}\ddot{y}_{1}+m_{2}\ddot{y}_{2}\right]\dot{u}dt.
	\label{NetWork2}
	\end{equation}
The two mass case above becomes clear if we first solve the single mass case, by letting $m_1=1$ and $m_2=0$ (see Sec.~\ref{dimen}). The average power input then is
	\begin{equation}
	W_{\rm single\, mass}=\frac{1}{T}\intop_{0}^{T}\ddot{y}_{1}\dot{u}dt=\frac{1}{T}\dot{y}_1\dot{u}\bigg|_{0}^{T}-\frac{1}{T}\intop_{0}^{T}\dot{y}_1\ddot{u}dt.
	\label{NetWork3}
	\end{equation}
	The term $\displaystyle \frac{1}{T}\dot{y}_1\dot{u}\bigg|_{0}^{T}$ in Eq.~(\ref{NetWork3}) goes to zero for large $T$ provided the energy harvesting device has a steady state behavior in which $\dot y_1$ remains uniformly bounded for all time\footnote{%
	Other arguments are possible. We could note that $|y_1|$ is
	bounded by $L/2$ (with nondimensionalization, $L=1$), and offers an infinite sequence of turning points; and we could choose a sequence of 
	$T$ values to 	
	coincide with those turning points.}.
Alternatively, in a simpler case, if $y_1$ is $2 \pi$-periodic like $u$, then choosing $T=2 \pi$ would make the average exactly zero. Either way, we can drop this term.

Substituting $u(t)=A\sin(t)$ in the second term, we have
	\begin{equation}
	W_{\rm single\, mass}=\frac{A}{T}\intop_{0}^{T} \sin(t)\dot{y}_1dt = \frac{A}{T} \sin(t) y_1 \bigg|_{0}^{T} -
\frac{A}{T}\intop_{0}^{T} \cos(t) y_1dt.
	\label{nonenergy2}
	\end{equation}
The first term on the right hand side is zero if $T=2n\pi$ for integer $n$. We are therefore left with 
$$W_{\rm single\, mass}= - \frac{A}{T}\intop_{0}^{T} \cos(t) y_1dt,$$
which achieves its upper bound if we allow $y_1$ (bounded in magnitude by 1/2) to be the discontinuous function
$$y_1 = - \frac{1}{2} \, {\rm sign } (\cos(t)).$$
The above limiting $y_1$ shows two things: (i) $y_1$ is $2 \pi$-periodic, and (ii) the upper bound
is
\begin{equation} \label{SMUB} W_{\rm single\, mass} \le  \frac{A}{\pi}.
\end{equation}

\begin{figure}[h!]
		\centerline{{\includegraphics[width=0.5\textwidth]{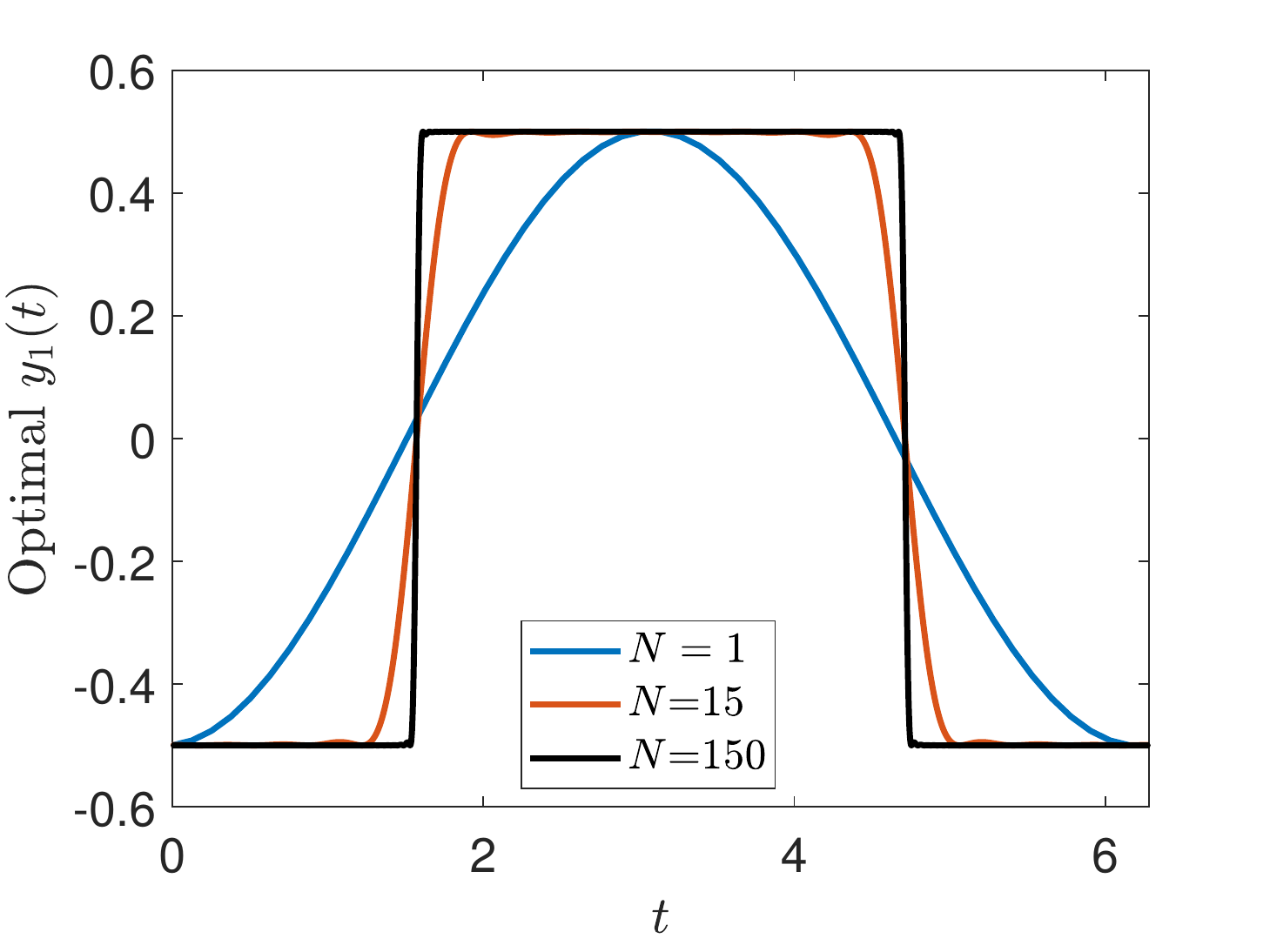}}}
		\caption{Optimal $y_1(t)$ for different $N$.}
		\label{figureSol}
	\end{figure}

Since the limiting $y_1$ is discontinuous (a physical impossibility for nonzero mass),
we study briefly the approach to the optimum using constrained Fourier series for a $2\pi$-periodic $y_1$, with numerical optimization.
Assuming
	\begin{equation}
	y_1(t)=\sum_{k=1}^{N}a_{k}\sin(kt)+b_{k}\cos(kt),
	\label{sersol}
	\end{equation}
	and substituting $y_1(t)$ in Eq.~(\ref{nonenergy2}), we obtain
	\begin{equation}
	W_{\rm single\, mass}=\frac{A}{2\pi}\intop_{0}^{2\pi}\sin(t)\dot{y}_1dt=-\frac{Ab_{1}}{2}.
	\label{nonenergy}
	\end{equation}
	Therefore, the least upper bound on the power extraction is obtained by finding the least upper bound of $-b_1$. In Eq.\ (\ref{sersol}) the bound $|y_1(t)| \le \frac{1}{2}$ (for $L=1$) restricts how large $-b_1$ can be. For finite $N$, the
problem can be approached numerically using linear programming as follows.
		We seek coefficients $a_k$ and $b_k$ in $y_1(t)$ (see Eq.~(\ref{sersol})), that maximize $-b_1$ while satisfying
	\begin{equation}
	\left|\sum_{k=1}^{N}a_{k}\sin(kt)+b_{k}\cos(kt)\right|\le\frac{1}{2} \quad \forall \, t.
	\label{constraints}
	\end{equation}
	By discretizing $t$ into $M$($\gg N$) equally spaced points between $0$ and $2\pi$, we can numerically obtain optimal solutions using, e.g., MATLAB's built-in function ``linprog''. Figure~\ref{figureSol} shows the optimal $y_1(t)$ for $N=1$, $N=15$, and $N=150$. It is clear that the optimal solution approaches a square wave as $N$ increases. From the Fourier series of a square wave, we have $b_{1}=-\frac{2}{\pi}$, and  substituting this value of $b_1$ in Eq.~(\ref{nonenergy}), the least upper bound on the (nondimensionalized)  power is found to match Eq.\ (\ref{SMUB}), which we now write as (recall Eq.\ (\ref{Wdef1}))
	\begin{equation}
	f_{1,\, \rm max,\, single\, mass} =\frac{A}{\pi}.
	\label{optimnon}
	\end{equation}
It follows that for the physical system in dimensional terms, the maximum\footnote{%
		Technically, it is a supremum and not a maximum; the term ``maximum'' is used informally. In practical terms, even 30\% of this value will be seen to be excellent.}
	possible power that can be extracted is
	$$ W_{\rm max,\, single\, mass} = \frac{m L A \omega^3}{\pi}.$$
We note that for an energy harvester whose dynamics is linear, with the response $y_1(t)$ containing only the first harmonic, the maximum possible energy output corresponds to $b_1 = -\frac{1}{2}$ with all other Fourier coefficients identically zero (see also the numerical result for $N=1$ in Fig.\ \ref{figureSol}), and in such cases
	$$ W_{\rm max, \, single\, mass,\, simple \, harmonic} = \frac{m L A \omega^3}{4},$$
i.e., the $\pi$ in the denominator is replaced by 4. The same proportion was observed in optimal motions of a whirling pendulum with gravity in~\cite{nandakumar2012optimum}.
	
\begin{figure}[h!]
		\centerline{{\includegraphics[width=0.5\textwidth]{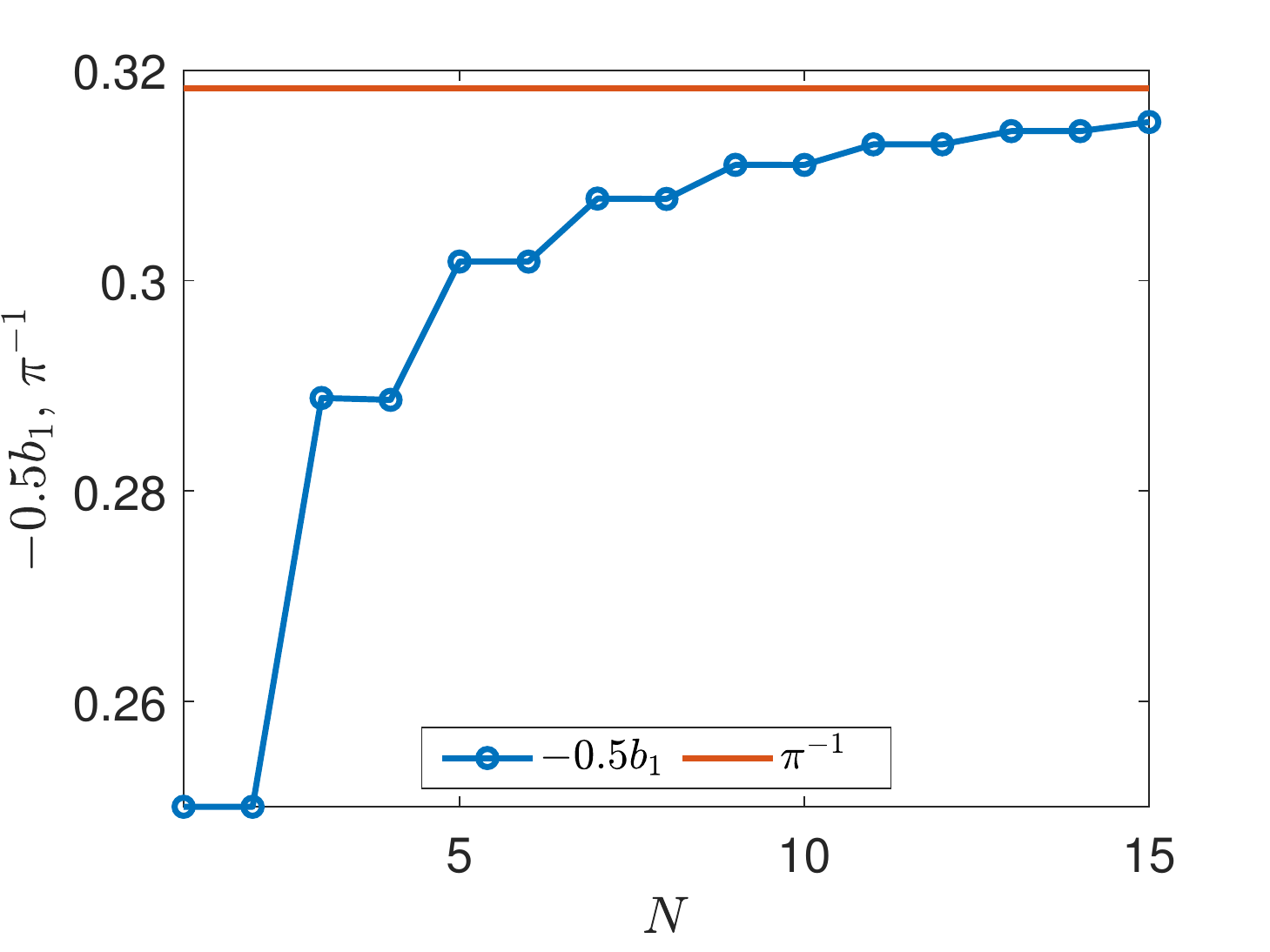}}}
		\caption{Convergence of $-0.5 \, b_1$ subject to Eq.\ (\ref{constraints}), with increasing $N$.}
		\label{convf}
	\end{figure}

	Figure~\ref{convf} shows the maximum achievable $f_{1,\, \rm single\, mass}$ for different $N$, obtained using linear programming. $A$ has been divided out, and $N$ varies from 1 to 15. While the limiting square wave displacement is not physically achievable, for $N=15$, 99\% of the theoretical maximum given in
	Eq.\ (\ref{optimnon}) is achieved. Hosseinloo and Turitsyn~\cite{hosseinloo2015non} proposed a theoretical design of an adaptive bistable harvester that can closely follow the square wave optimal solution, but it was not realized experimentally.

With the above insights, we can temporarily drop the $2 \pi$-periodicity assumption on both $u$ and $y_1$, and average over a long time $T$. Using arguments similar to those used above, we find 
	\begin{equation}
	f_{1,\, \rm max,\,single\, mass,\, aperiodic}=\lim_{T \rightarrow \infty} \frac{1}{2T}\intop_{0}^{T}\left|\frac{d^{3}u}{dt^{3}}\right|dt,
	\label{NetWork61}
	\end{equation}
which reduces to expressions obtained above for the purely sinusoidal case. For the rest of this paper, we consider $u = A \sin ( \omega t)$.

We now return to the two-mass case, Eq.\ (\ref{NetWork2}), with $m_1+m_2 = 1$. Here it helps to note an almost obvious fact from optimization theory. Let $F(\xi)$ be a general scalar function of a general vector variable $\xi$. Let us consider two generic optimization problems, with a hierarchical relationship as follows.

\bigskip
	
	\noindent {\bf Problem 2:} Maximize $F(\xi)$ subject to the constraints $C_1(\xi)=0$ and $C_2(\xi) \ge 0$.

	\noindent {\bf Problem 3:} Maximize $F(\xi)$ subject to the constraints $C_1(\xi)=0$ and $C_2(\xi) \ge 0$, along with additional constraints $C_3(\xi)=0$ and/or $C_4(\xi) \ge 0$.

\bigskip
	
In the above two problems, the objective functions are the same. Every constraint of problem 2 applies to problem
3 as well. The only difference is that problem 3 has additional constraints.
Now, suppose problem 2 has a solution with a finite maximum, say $F_2$ for $\xi=\xi_2$. Suppose further that problem 3 has a nonempty feasible set, i.e., there is at least one point $\xi$ which satisfies all the constraints of problem 3; then that $\xi$ automatically satisfies all the constraints of problem 2 as well. More importantly, problem 3 also has a solution with a finite maximum, say $F_3 \le F_2$ at $\xi=\xi_3$ (if $F_3 > F_2$ then we have a contradiction). Finally, if the maximizer $\xi_2$ of problem 2 satisfies the constraints of problem 3, then $\xi_2$ is a maximizer of
problem 3 as well, and $F_3 = F_2$.

In other words, if we already have a solution to some optimization problem, and then we add on some more constraints and solve the optimization problem again, the optimum {\em does not improve}. In the special case where the solution to the original problem satisfies the added constraints of the new problem, the original optimizer provides a solution for the new problem as well.

With the above simple insight, we can revisit the expression given by Eq.~(\ref{NetWork2}) for the work done on a device with two point masses. Considering $m_1=\gamma $ and $m_2=1-\gamma$ (i.e., $m_1+m_2=1$), we have 
	\begin{equation}
	f_{1,\rm \thinspace two\thinspace masses}=\frac{\gamma}{T}\intop_{0}^{T}\ddot{y}_{1}\dot{u}dt+\frac{1-\gamma}{T}\intop_{0}^{T}\ddot{y}_{2}\dot{u}dt.
	\label{NetWork7}
	\end{equation}
The design of the device may make $y_1$ and $y_2$ interrelated, i.e., there may be further constraints on them in addition to each being bounded in magnitude by $L/2$ (or $1/2$, with nondimensionalization). With those further constraints, this problem corresponds to problem 3 above. However, we can first ignore those further constraints, which reduces to problem 2 above. That problem, with
$y_1$ and $y_2$ independent, has already been solved above (note that the mass was taken above as unity, and now needs to be reintroduced for each term):
	\begin{equation}
	f_{1,\,\rm max,\thinspace two\thinspace masses}=\frac{\gamma A}{\pi}+\frac{(1-\gamma)A}{\pi}=\frac{A}{\pi}.
	\label{NetWork8}
	\end{equation}
Although we do not know the constraints that different designs may place on the motions of masses $m_1$ and $m_2$, we know the
upper bound above, of $A/\pi$, cannot be crossed (this corresponds to $F_3 \le F_2$ in the discussion above). Additionally, of all those possible sets of constraints, we must allow two cases: (i) the constraint $y_1=y_2$, which means a single point mass, and (ii) no constraints on $y_1$ and $y_2$,
e.g., two uncoupled oscillators placed side by side within the same casing. In each case, the upper bound of $A/\pi$ is achieved, and therefore it cannot be lowered (this corresponds to $F_3 = F_2$ above).

It follows by similar reasoning that for a system comprising $p$ point masses $m_k=\gamma_k,\,k=1,2,..,p$, with $\displaystyle \sum_{k=1}^{p}\gamma_{k}=1$, the upper bound remains the same:
$$f_{1,\,\rm max,\thinspace multiple\thinspace point \thinspace masses}=\frac{A}{\pi}.$$

It remains to consider energy harvesting devices that have bodies whose mass is not concentrated into point masses. There are two situations to consider: (i) rigid bodies, and (ii) flexible bodies.

	\begin{figure}[h!]
	\centerline{{\includegraphics[width=0.8\textwidth]{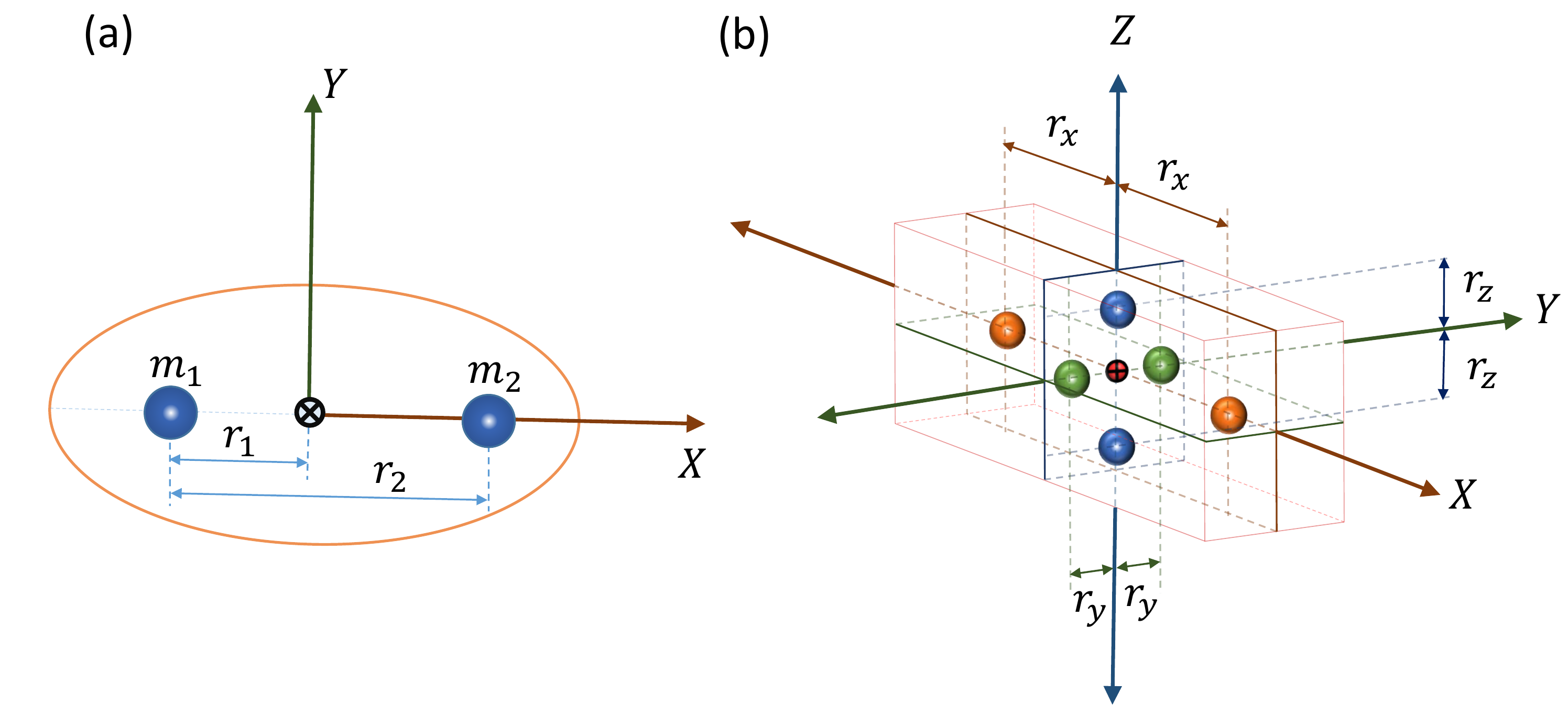}}}
	\caption{Equivalent representation of (a) planar and (b) spatial rigid bodies using point masses. For dynamical equivalence, a planar rigid body with mass $m$ and moment of inertia $I$ must satisfy, $m_1+m_2=m$, $m_1 r_1= m_2 (r_2-r_1)$, and $I=m_1 r_1^2+m_2 (r_2-r_1)^2$. The spatial rigid body with mass $m$ is represented using six equal masses of $\frac{m}{6}$, located symmetrically about the center of mass along each principal axis at distances $r_x$, $r_y$, and $r_z$ as shown. The distances must be selected such that $I_{xx}=\frac{m}{3}(r_y^2+r_z^2)$, $I_{yy}=\frac{m}{3}(r_x^2+r_z^2)$, and $I_{zz}=\frac{m}{3}(r_x^2+r_y^2)$. }
	\label{3D2D}
\end{figure}

Rigid bodies are simple, and planar motion is simpler than spatial motion.

In planar motion (see Fig.~\ref{3D2D}a), any rigid body is dynamically equivalent to a rigidly attached pair of point masses, say $m_1$ and $m_2$, separated by some distance $r_2$, such that the center of mass lies at some specified point on the line joining the two point masses (i.e., we specify $r_1$). These three quantities, $m_1$, $m_2$ and $r_2$, must be assigned values such that the total mass, the center of mass location, and the moment of inertia about the center of mass of this equivalent body match those of the original body (three equations in three unknowns). Subsequently, each such rigid body becomes equivalent to a pair of point masses with an added constraint that the distance between them does not change, which is the situation of problem 2 above being restricted to problem 3. By foregoing arguments, therefore, the upper bound for energy extracted cannot increase due to the presence of rigid bodies in planar motion.

Three dimensional rigid bodies undergoing spatial motions  are also dynamically equivalent (see Fig.~\ref{3D2D}b) to a finite number of point masses connected rigidly together. 
One simple approach is to first locate the center of mass and compute the principal axes of inertia.
Then, place 6 point masses, each equal to 1/6 of the total mass, pairwise and symmetrically about the center of mass and on the three principal
axes. Three distances from the origin remain to be chosen, to
match three principal moments of inertia. Real solutions exist if the principal moments of inertia 
satisfy three inequalities of the form $I_{ii} \le I_{jj} + I_{kk}$, which hold for real rigid bodies.
Subsequently, the same arguments regarding added constraints apply, and the bound remains the same:
$$f_{1,\,\rm max,\thinspace with\thinspace rigid \thinspace bodies}=\frac{A}{\pi}.$$

\begin{figure}[h!]
		\centerline{{\includegraphics[width=0.5\textwidth]{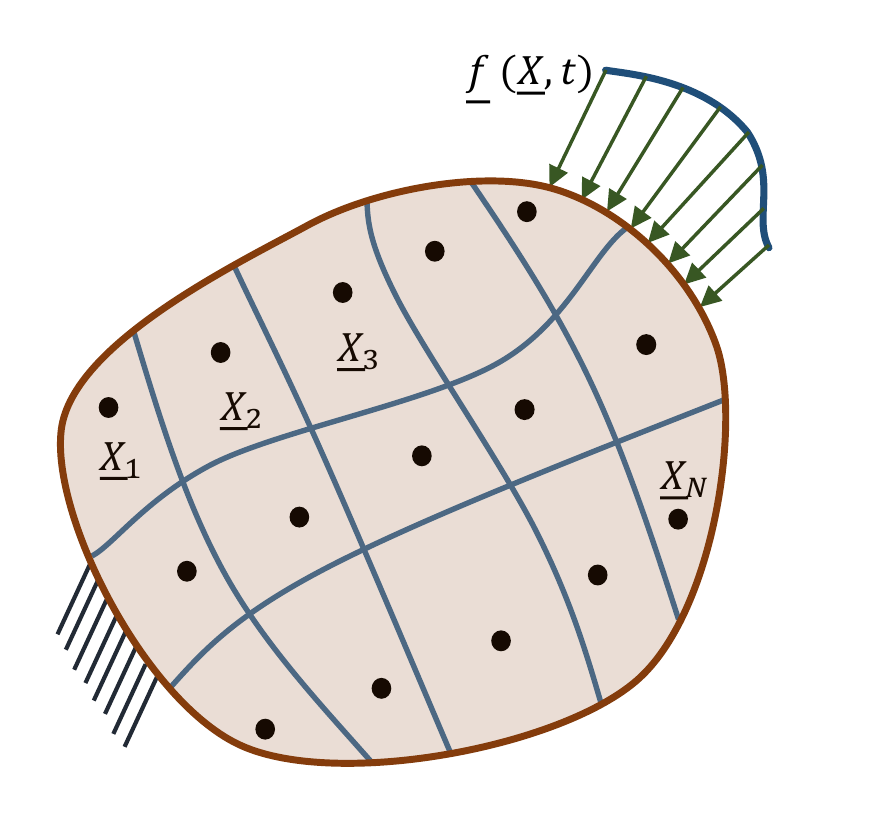}}}
		\caption{An arbitrary body.}
		\label{mesh}
	\end{figure}

Next, we turn to deformable or flexible bodies with distributed mass. See Fig.\ \ref{mesh}, which shows a schematic of a single arbitrary body. A portion of the boundary is shown fixed, representing some essential boundary conditions . Under base excitation, those boundary points will be given a displacement $u(t) \hat j$. Also, distributed forces $\underline f({\bf X},t)$ can act. The body is shown divided into a finite number of subregions or elements, for reasons that will be clear below.

The undeformed three-dimensional domain is 
$\Omega_0$, wherein material points are located using reference position vectors ${\bf X}$. The undeformed density is $\rho_0({\bf X})$. Subsequently, under dynamic motions, the displaced points of the body are assumed to be accurately described by
\begin{equation}
\label{df1} {\bf x} = {\bf X} + u(t) \hat j + \sum_{k=1}^n q_k(t) \underline \phi_k({\bf X}),
\end{equation}
where $u(t) \hat j$ is the base excitation, the $\underline \phi_k$ represent kinematically admissible basis functions,  the $q_k(t)$ represent generalized coordinates, and $n$ modes give a sufficiently accurate solution for the purpose at hand.
After deformation, the displaced configuration is $\Omega$, wherein the density is $\rho({\bf x},t)$.

Consider writing Lagrange's equations for this body under base excitation. We will need the strain energy, which will involve integration of the strain energy density over the body and will depend nonlinearly on the $q_k(t)$; the details are not important here. Generalized forces $Q_k(t)$ will similarly be computed through integrals of virtual work density over the domain, and will possibly depend on the external forces $f$ as well as the deformation, the deformation rate, and perhaps even their time histories (if there is hysteresis in the response). The key point is that they will not depend explicitly on the density of the material. Finally, we will consider kinetic energy, which is the only place where the mass distribution plays a role.

The kinetic energy computed over the instantaneous or current configuration is
$${\rm KE} = \frac{1}{2} \int_{\Omega} \rho({\bf x},t) \, {\bf v} \! \cdot \! {\bf v} \, d {\rm Vol},$$
where ${\bf v}$ is the instantaneous absolute (or inertial) velocity of the material point that is instantaneously at spatial location ${\bf x}$, and $ d {\rm Vol}$ represents an infinitesimal volume element in the deformed body. The integral can also
be computed on the undeformed domain, by using
$$\rho({\bf x},t) \, d {\rm Vol} = \rho({\bf X}) \, d {\rm Vol}_0,$$
where $d {\rm Vol}_0$ is the volume that was occupied by element $d {\rm Vol}$ when it was in the reference configuration. Thus,
$${\rm KE} = \frac{1}{2} \int_{\Omega_0} \rho_0({\bf X}) \, {\bf v} \! \cdot \! {\bf v} \, d {\rm Vol}_0.$$
The integrand in the above is
$$ \rho_0({\bf X}) \left ( \dot u(t) \hat j + \sum_{k=1}^n \dot q_k(t) \underline \phi_k({\bf X}) \right ) \cdot
\left ( \dot u(t) \hat j + \sum_{k=1}^n \dot q_k(t) \underline \phi_k({\bf X}) \right ),$$
and the integration domain $\Omega_0$ is shown in Fig.\ \ref{mesh}. If we divide $\Omega_0$ into $N$ small elements, and choose one representative point each (say, ${\bf X}_1, {\bf X}_2, \cdots, {\bf X}_N$) in these elements, then the integral is well approximated by
\begin{equation}
\label{KEa}
KE \approx \frac{1}{2} \sum_{r=1}^N \rho_0({\bf X_r}) \, {\rm Vol}_r \, \left ( \dot u(t) \hat j + \sum_{k=1}^n \dot q_k(t) \underline \phi_k({\bf X_r}) \right ) \cdot
\left ( \dot u(t) \hat j + \sum_{k=1}^n \dot q_k(t) \underline \phi_k({\bf X_r}) \right ),
\end{equation}
where ${\rm Vol}_r$ is the volume of the $r^{\rm th}$ element.
With $m_r = \rho_0({\bf X_r}) \, {\rm Vol}_r,$
$$KE \approx \frac{1}{2} \sum_{r=1}^N m_r \, \left ( \dot u(t) \hat j + \sum_{k=1}^n \dot q_k(t) \underline \phi_k({\bf X_r}) \right ) \cdot
\left ( \dot u(t) \hat j + \sum_{k=1}^n \dot q_k(t) \underline \phi_k({\bf X_r}) \right ).$$
For large but finite $N$, the above is a good approximation; as $N \rightarrow \infty$ with each element shrinking to zero size, the integral converges. Note that $N$ may be much bigger than $n$.

Finally, imagine a hypothetical body that has the same shape, the same material properties except density, and the same displacement field, i.e., the motion constraints of Eq.~(\ref{df1}). The only difference is that this body is made of an otherwise massless material except for point masses $m_r$ embedded
at points $X_r$. It is clear that Lagrange's equations for this hypothetical body, and the original body as approximated by Eq.~(\ref{KEa}), are {\em identical}. The energy extractable, as per the two models, will be identical. Yet, the second one has a finite number of point masses with added motion constraints, and so the same upper bound applies for extractable power.

In contrast to the above detailed argument, we can use a more direct argument as well. Suppose that the energy harvester consists of a collection of possibly deformable bodies. These bodies together, at any configuration, have a total mass $m$ and a center of mass location $x_G \hat i + (u + y_G) \hat j + z_G \hat k$. Internal forces between these bodies and the casing do not appear in the free body diagram.
Linear momentum balance yields (compare with Eq.~(\ref{Fnet}))  
\begin{equation}
	\underline{F}_{\textrm{net} }= - \underline{Q} + m\ddot{x}_{G}\,\hat{\textrm{i}}+\left[m(\ddot{y}_{G}+\ddot{u})+M\ddot{u}\right]\hat{\textrm{j}} + m\ddot{z}_{G}\, \hat{\textrm{k}}.
	\nonumber
	\end{equation}
If every point in each moving body is motion limited within a range $L$, then so is $y_G$. The arguments following Eq.~(\ref{Fnet}) then apply, and we conclude that the upper bound on the extracted power is the same. This latter line of argument is preferable to some readers because it is shorter. Other readers may find it intuitively less satisfactory to consider motions of the single center of mass of a collection of arbitrarily moving and deforming bodies with arbitrary shapes. Such readers may like to consider the parallel and longer preceding argument.

In either case, the upper bound on extractable power remains the same.
	
\begin{figure}[htpb!]
		\centerline{\includegraphics[width=0.8\textwidth]{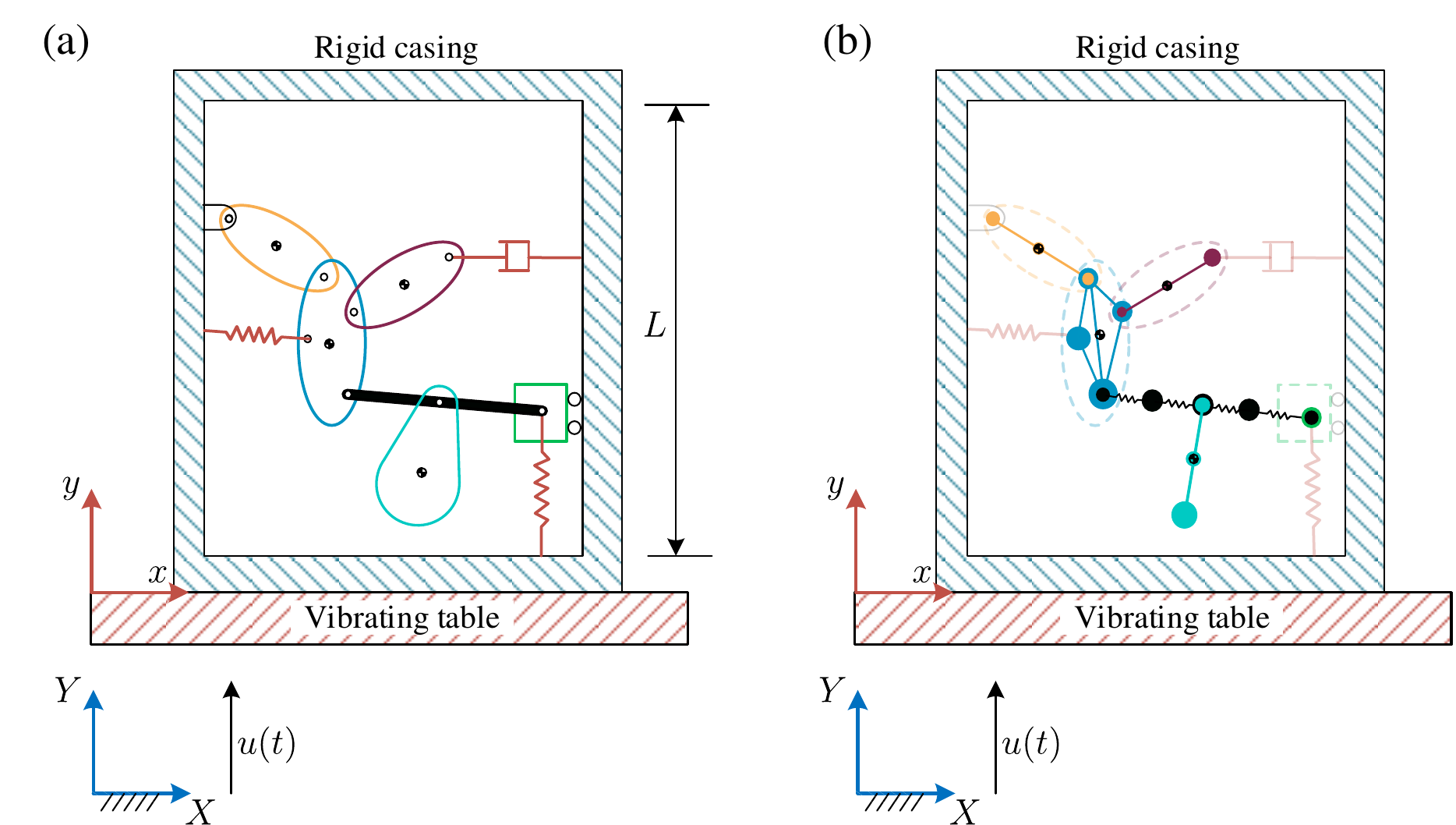}}
		\caption{A multibody energy harvester. (a) Physical system. (b) Equivalent system using point masses with motion constraints.}
		\label{Schematic4}
	\end{figure}

The ideas described above are illustrated schematically in Fig.\ \ref{Schematic4}.
In Fig.~\ref{Schematic4}(a), we show an arbitrary multibody energy harvester under harmonic base excitation. For simplicity of depiction, we assume that the black rod has axial vibrations and the other components are rigid. We assume further that a four-element model of the black rod is sufficiently accurate. 	
Figure~\ref{Schematic4}(b) shows an equivalent point mass representation of the same system. Each rigid body is represented as a dynamically equivalent set of point masses (the representation is nonunique). The axially-deformable rod, by the  point mass discussion above, is taken to be equivalent to five point masses connected by deformable elements with some motion constraints that depend on the type of element used. 
Our bound holds, and the maximum average power that can be extracted from the device is bounded by $\displaystyle \frac{mLA\omega^3}{\pi}$.
	
To summarize, under harmonic base excitation, irrespective of the internal construction of the energy harvester, if its total internal moving mass is $m$, and if no material point in the device is allowed to have displacements that exceed a total magnitude $L$, then the maximum average power that can be transmitted by the moving base on the casing is $\displaystyle \frac{mLA\omega^3}{\pi}$.

Although we have a general upper bound on power that can be extracted by the energy harvester, we have not established what fraction of that limit a practical device may be able to extract.  We turn to two such special cases next.
	
	\section{Linear resonant oscillator with an ideal generator}
	\label{linres}
Williams and Yates~\cite{williams1996analysis} found an expression for the maximum power that can be extracted from a linear spring-mass oscillator. Here, we state and slightly extend the results for completeness and later discussion.

Suppose we have a linear base-excited oscillator such that the power-extracting device exerts a resistive force proportional to relative velocity ($-\zeta \dot{y}$).  Other damping is assumed negligible. The equation of motion is
	\begin{equation}
	m\ddot{y}+\zeta\dot{y}+ky=mA\omega^{2}\sin(\omega\tau).
	\label{lineq}
	\end{equation}
	By substituting $m=1$, $L=1$, and $\omega=1$ (see our nondimensionalization strategy in Sec.~\ref{dimen}), and setting $k=1$ for resonance under light damping, we obtain
	\begin{equation}
	\ddot{y}+\zeta\dot{y}+y=A\sin(t).
	\label{linnd}
	\end{equation}
	The steady state solution of Eq.~(\ref{linnd}) is
	\begin{equation}
	y(t)=-\frac{A}{\zeta}\cos(t).
	\end{equation}
	The  power extracted is
	\begin{equation}
	f_{1,\rm  \, linear \,  resonator}=\frac{1}{2\pi}\int_{0}^{2\pi}\zeta\dot{y}^{2}dt=\frac{A^{2}}{2\pi\zeta}\int_{0}^{2\pi}\sin^{2}tdt=\frac{A^{2}}{2\zeta}.
	\label{LRener}
	\end{equation}
	Here, $\zeta$ cannot be decreased below some limit. The maximum allowable amplitude for $y(t)$ is $0.5$ (design constraint). Therefore,
	\begin{equation}
	\left|y\right|=\frac{A}{\zeta} \le \frac{1}{2}, \mbox{ or } \zeta \ge 2A.
	\end{equation}
	Maximum power output occurs for $\zeta=2A$,
	\begin{equation} \label{mpe}
	f_{1,\,\rm max,\thinspace linear\thinspace resonator} =\frac{A}{4}.
	\end{equation}
	The above is  21.5\% less than the theoretical upper bound ($f_{1,\, \rm max,\, single\, mass} =\frac{A}{\pi}$),  and agrees with the solution obtained with one Fourier term ($N=1$) in Section~\ref{Nonlinaer}, see Fig.\ \ref{figureSol}.

It is interesting to consider the case where the inherent damping in the system is not negligible, with
$$\zeta = \zeta_m + \zeta_e,$$
where $\zeta_m$ is already-present mechanical damping, and $\zeta_e$ is the additional effect of the energy extraction device. We now have the amplitude constraint
\begin{equation} \label{opz} \zeta_e + \zeta_m \ge 2A, \end{equation}
while the power extracted works out to
$$\frac{\zeta_e A^2}{2 (\zeta_m+ \zeta_e)^2}.$$
Now the maximum power extracted is less than that in Eq.\ \ref{mpe}. In particular, if $\zeta_m \ge A$, the
maximum power extracted equals 
$$\frac{A^2}{8 \zeta_m} \mbox{ for } \zeta_e = \zeta_m.$$
The above expression explains what happens if $\zeta_m$ is fixed and $A$ becomes small: the power extracted becomes quadratic instead of linear in $A$, leading to low conversion efficiencies.

	\section{Whirling pendulum with linearly proportional generator torque}
	\begin{figure}[h!]
		\centerline{{\includegraphics[width=0.35\textwidth]{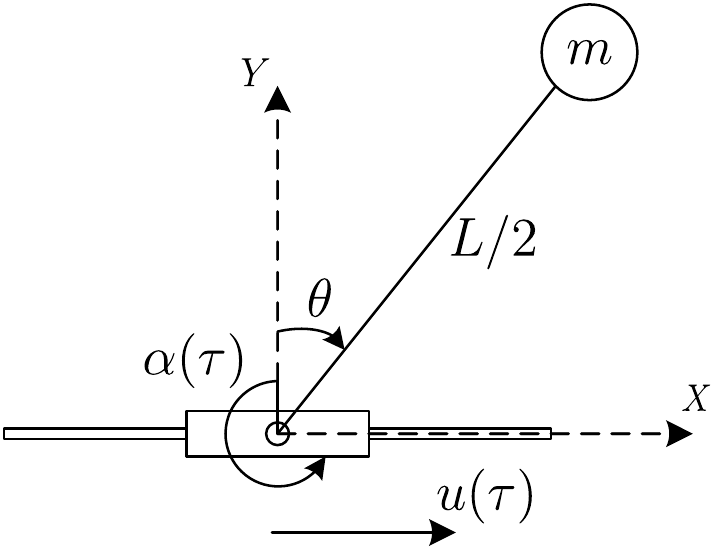}}}
		\caption{Schematic of a whirling pendulum with an ideal generator.}
		\label{Schematic_pen}
	\end{figure}
As mentioned in the introduction, there are pendulum based wave energy harvesting applications where gravity plays an important role. However, if the pendulum moves in a horizontal plane then gravity plays no role. An advantage of such a device is that it does not have a preferred resonant frequency: the pendulum can whirl at any speed.
	In this section we study a base-excited whirling pendulum, as shown in Fig.~\ref{Schematic_pen}. We assume the base excitation $u(t)=A\sin(\omega t)$ and the resisting torque is $-\zeta \dot{\theta}$, due entirely to an ideal generator which converts mechanical work inputs fully into electrical power. The pendulum is assumed to have a single point mass $m$, and
a massless rigid rod of length $L/2$. Its motion is governed by
	\begin{equation}
	\frac{mL^{2}}{4}\ddot{\theta}+\zeta\dot{\theta}-\frac{m\omega^{2}AL}{2}\sin(\omega t)\sin(\theta)=0.
	\label{pendeq}
	\end{equation}
	Following our nondimensionalization strategy of Sec.~\ref{dimen}, we substitute $m=1$, $L=1$, and $\omega=1$ in Eq.~(\ref{pendeq}), obtaining
	\begin{equation}
	\ddot{\theta}+4\zeta\dot{\theta}-2A\sin(t)\sin(\theta)=0.
	\label{pendeqnd}
	\end{equation}
	
	\begin{figure}[h!]
		\centerline{{\includegraphics[width=0.8\textwidth]{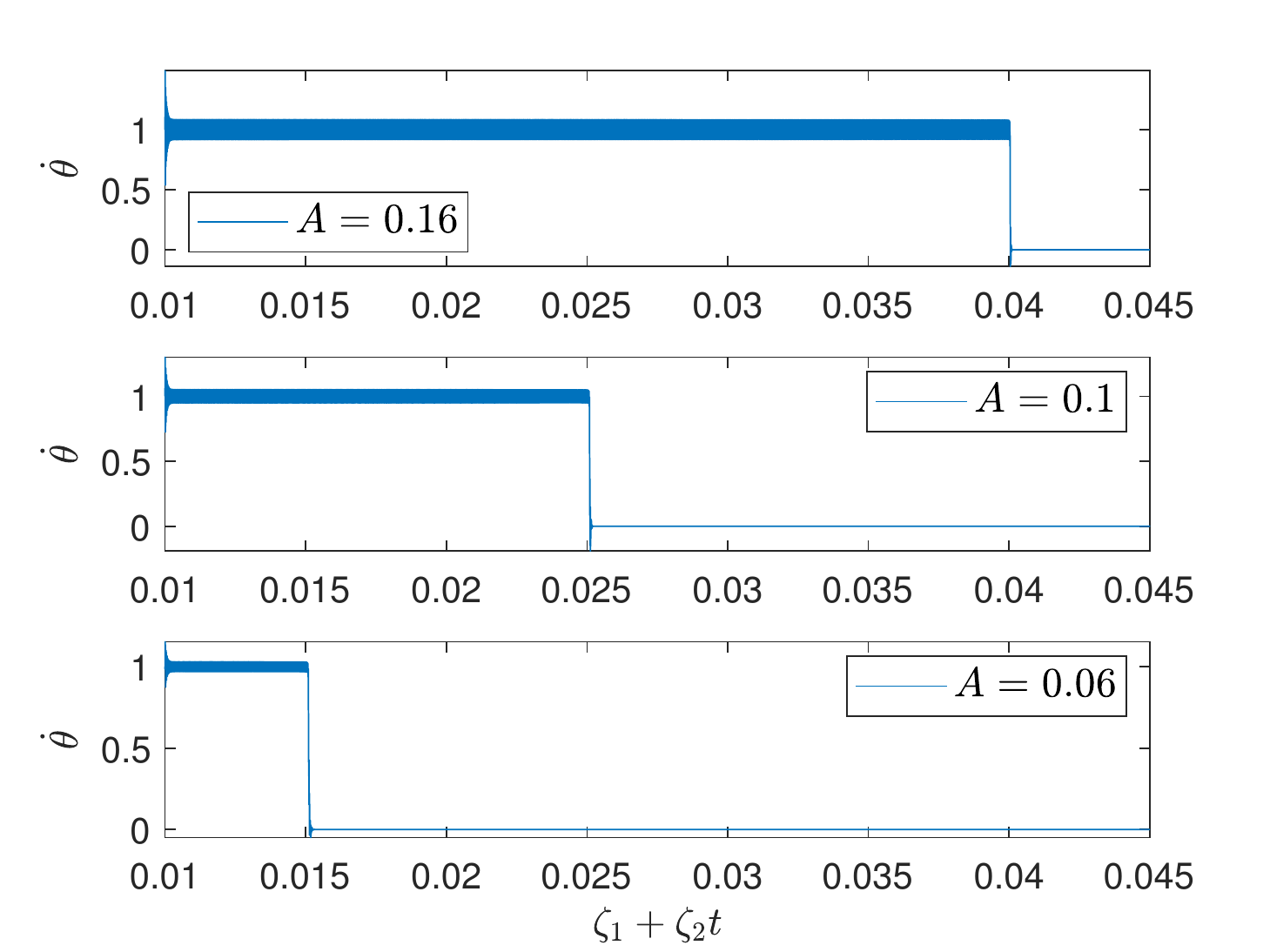}}}
		\caption{Numerical investigation of synchronous whirling solutions for slowly increasing $\zeta$ and different values of $A$. $\dot \theta$ is plotted against $\zeta$ in each subplot. }
		\label{RPS}
	\end{figure}
	Now, we ask the following question: for synchronous whirling motion, i.e. for an average rate of $\dot{\theta}=1$, what is the maximum  power that can be extracted? In other words, what is the maximum value of $\zeta$ that can be used without quenching the synchronous whirling motion?  To answer this question, we first use harmonic balance~\cite{nayfeh2008applied} and seek a solution of the form
	\begin{equation}
	\theta = t + \theta_0.
	\label{solpen}
	\end{equation}
	Equation (\ref{pendeqnd}) yields
	\begin{equation}
	4\zeta=-2A\sin(t)\left(\sin(t)\cos(\theta_{0})+\cos(t)\sin(\theta_{0})\right).
	\label{harm}
	\end{equation}
	The constant on the left side of Eq.~(\ref{harm}) is matched with the average of the right side, yielding
	\begin{equation}
	\cos(\theta_0)=\frac{4\zeta}{A}.
	\label{damping}
	\end{equation}
	Equation~(\ref{damping}) has no solution when $\left|\frac{4\zeta}{A}\right|>1$. However, if $\left|\frac{4\zeta}{A}\right|<1$, it has two symmetric solutions corresponding to the positive and negative values of $\theta_0$. In the limiting case,  $\frac{4\zeta}{A}=1$. Hence, the maximum power is extracted when $\zeta=\frac{A}{4}$.  The power that can be extracted from the pendulum  (see Eq.~(\ref{pendeq})) is
	\begin{equation}
	f_{\rm 1, \, ideal\, generator}=\zeta\frac{1}{2\pi}\intop_{0}^{2\pi}\left(\frac{d\theta}{dt}\right)^{2}dt.
	\end{equation}
	Since $\dot{\theta}=1$, the dimensionless power is equal to $\zeta$ and attains a maximum at $\zeta=\frac{A}{4}$, i.e.
	$$f_{\rm 1,\,max, \, ideal\, generator} =\frac{A}{4}.$$
This value of maximum dimensionless power is exactly the same as the maximum  power extractable from the resonant linear oscillator  in Section~\ref{linres}. This rotary design has the advantage of not having a preferred frequency, and no significant deformation. The translating resonant oscillator design has the advantage of not needing a hinge joint.

The foregoing approximate analysis assumes that the solution $\theta = t + \theta_0$ is stable. Using numerical simulations, we can
verify easily that one synchronous whirl solution is stable when $0 < \zeta \lesssim \frac{A}{4}$. To demonstrate for any given value of $A$, we can increase the damping slowly with time within a simulation, using $\zeta=\zeta_0+\zeta_1 t$, and see how far the whirling solution survives. 
	
Figure~\ref{RPS} shows numerical solutions of Eq.~(\ref{pendeqnd}) with initial conditions $\theta(0)=0$ and $\dot{\theta}(0)=1$. In Fig.~\ref{RPS}, $\dot \theta$ is plotted against $\zeta_0+\zeta_1t$ with $\zeta_0=1\times 10^{-2}$ and $\zeta_1=1\times 10^{-6}$. It is clear that the synchronous whirling solution is stable until $\zeta$ becomes equal to $\frac{A}{4}$, for different $A$.
	
We conclude that, in terms of extractable power, a linear resonant oscillator and a whirling pendulum with a resisting torque proportional to whirl rate have the same upper bound, which is $\pi/4$ times the theoretical upper bound for arbitrary nonlinear multi-degree-of-freedom designs.

This concludes the theoretical discussion of the paper. We now have a simple upper bound on the power extractable by an energy harvester. We can now examine several devices reported in the literature, and report the percentage of the theoretical maximum they were actually able to extract.
	
	\section{Performance of some existing devices}
	\label{Performance}
In this section, we examine several published works and compare the actual power extracted, $W_{\textrm{exp}}$ with the theoretical upper bound for that device, $W_{\textrm{max}}$. In Table~\ref{table}, the efficiency
$$\eta=\frac{W_{\textrm{exp}}}{W_{\textrm{max}}}\times 100\%,$$
is reported from 36 studies. These efficiencies range from $0.0036$\% to $29$\%. The theoretical maximum power extractable for these devices, based on their dimensional physical parameters, ranged from under a milliwatt to 30 watts. The actual power extracted ranged from a few microwatts to several watts.
All the devices listed in Table~\ref{table} are oscillatory by design except \cite{smilek2019rolling,sasaki2005vibration} where full whirling motion is reported.

	\begin{table}[htpb!]
			\centering
		\caption{Efficiencies of some physically realized energy harvesting devices from the literature.}
		\begin{tabular}{|c|c|c|c|c|c|c|c|}
			\hline
			No.\ [Ref.] & $m$ (kg) & $L$ (m) & $\omega$ (s$^{-1}$) & $A\omega^2$  (m s$^{-2}$) & $W_{\textrm{exp}}$ (W) & $W_{\textrm{max}}=\frac{mLA\omega^3}{\pi}$ (W) & $\eta$ \\ \hline
			1.\ \cite{renaud2009harvesting} & $6.00\times10^{-2}$  & $3.50\times10^{-2}$  & $6.28\times 10^{1}$ & $3.94\times 10^{2}$  & $6.00\times10^{-4}$ & $1.66\times 10^{1}$ & $0.0036\%$ \\ \hline					
			2.\ \cite{moss2013wideband} & $6.70\times10^{-2}$  & $7.00\times10^{-3}$  & $5.02\times 10^{1}$ & $1.70$  & $3.30\times10^{-6}$ & $1.32\times10^{-2}$ & $0.025\%$ \\ \hline
			3.\ \cite{ju2013low} & $4.91\times10^{-4}$  & $2.50\times10^{-3}$  & $1.07\times 10^{2}$ & $2.94\times10^1$  & $3.60\times10^{-7}$ & $1.23\times10^{-3}$ & $0.029\%$ \\ \hline
			4.\ \cite{marszal2017energy}	&  $5.00$ & $0.23$  & $8.90\times 10^{1}$ & $3.96\times 10^{1}$  & $4.00\times 10^{-1}$ & $1.29e3$ & $0.031\%$ \\ \hline
			5.\ \cite{miki2010large} & $1.10\times10^{-5}$  & $1.65\times10^{-2}$  & $3.96\times 10^{2}$ & $1.96\times10^1$  & $1.00\times10^{-6}$ & $4.50\times10^{-4}$ & $0.222\%$ \\ \hline
			6.\ \cite{halim2015modeling} & $4.36\times10^{-3}$  & $4.00\times10^{-2}$  & $3.64\times 10^{1}$ & $1.96\times 10^{1}$  & $1.03\times10^{-4}$ & $3.97\times 10^{-2}$ & $0.260\%$ \\ \hline	
			7.\ \cite{ashraf2013wideband} & $4.70\times10^{-2}$  & $6.00\times10^{-2}$  & $6.28\times 10^{1}$ & $9.81$  & $1.80\times10^{-3}$ & $5.53\times 10^{-1}$ & $0.326\%$ \\ \hline
			8.\ \cite{galchev2011micro} & $9.30\times10^{-3}$  & $2.70\times10^{-2}$  & $6.28\times 10^{1}$ & $9.81$  & $1.63\times10^{-4}$ & $4.92\times 10^{-2}$ & $0.331\%$ \\ \hline
			9.\ \cite{liu2012new} & $3.08\times10^{-6}$  & $3.40\times10^{-3}$  & $1.72\times 10^{2}$ & $5.89\times 10^{-1}$  & $1.12\times10^{-9}$ & $3.36\times 10^{-7}$ & $0.331\%$ \\ \hline
			10.\ \cite{galchev2012piezoelectric} & $9.30\times10^{-3}$  & $1.40\times10^{-2}$  & $6.28\times 10^{1}$ & $9.81$  & $1.00\times10^{-4}$ & $2.55\times 10^{-2}$ & $0.392\%$ \\ \hline
			11.\ \cite{huang2007silicon} & $7.50\times10^{-5}$  & $1.00\times10^{-3}$  & $6.28\times 10^{2}$ & $1.97\times 10^{1}$  & $1.44\times10^{-6}$ & $2.96\times10^{-4}$ & $0.486\%$ \\ \hline
			12.\ \cite{tang2011bi} & $4.60\times10^{-3}$  & $2.20\times10^{-2}$  & $6.28\times 10^{1}$ & $9.81$  & $1.01\times10^{-4}$ & $1.98\times10^{-2}$ & $0.511\%$ \\ \hline
			13.\ \cite{bendame2016wideband} & $1.21\times 10^{-1}$  & $6.00\times10^{-2}$  & $1.16\times 10^{2}$  & $5.90$ & $1.20\times10^{-2}$  & $1.57$ & $0.764\%$  \\ \hline
			14.\ \cite{dai2016vibration} & $8.20\times10^{-2}$  & $9.50\times10^{-3}$  & $9.31\times 10^{1}$  & $4.91$ & $9.70\times10^{-4}$  & $1.13\times10^{-1}$ & $0.858\%$  \\ \hline
			15.\ \cite{sasaki2005vibration} & $4.70\times10^{-3}$  & $2.70\times10^{-2}$  & $1.25\times 10^{1}$ & $3.92\times 10^{1}$  & $1.71\times10^{-4}$ & $1.98\times10^{-2}$ & $0.864\%$ \\ \hline
			16.\ \cite{liu2015wideband} & $9.00\times10^{-3}$  & $3.90\times10^{-2}$  & $4.15\times 10^{2}$ & $6.00$  & $2.64\times10^{-3}$ & $2.78\times10^{-1}$ & $0.950\%$ \\ \hline
			17.\ \cite{malaji2015analysis} & $2.60\times10^{-2}$  & $2.70\times 10^{-1}$  & $1.30\times 10^{1}$ & $3.40\times 10^{-1}$  & $1.10\times10^{-4}$ & $1.04\times10^{-2}$ & $1.058\%$ \\ \hline	
			18.\ \cite{salauddin2016magnetic} & $1.21\times10^{-2}$  & $7.00\times10^{-2}$  & $6.91\times 10^{1}$ & $4.91$  & $1.09\times10^{-3}$ & $9.14\times10^{-2}$ & $1.196\%$ \\ \hline
			19.\ \cite{wang2017non} & $2.50\times10^{-3}$  & $3.00\times10^{-3}$  & $1.57\times 10^{2}$ & $1.96\times 10^{1}$  & $9.00\times10^{-5}$ & $7.35\times10^{-3}$ & $1.224\%$ \\ \hline
			20.\ \cite{moss2012bi} & $6.70\times10^{-2}$  & $9.00\times10^{-3}$  & $6.09\times 10^{1}$ & $6.00\times 10^{-1}$  & $1.21\times10^{-4}$ & $7.07\times10^{-3}$ & $1.712\%$ \\ \hline
			21.\ \cite{smilek2019rolling} & $5.60\times10^{-2}$  & $5.00\times10^{-2}$  & $1.75\times 10^{1}$ & $1.27\times 10^{1}$  & $5.00\times10^{-3}$ & $1.98\times10^{-1}$ & $2.525\%$ \\ \hline
			22.\ \cite{malaji2017magneto} & $2.60\times10^{-2}$  & $1.75\times10^{-1}$  & $1.50\times 10^{1}$ & $5.70\times 10^{-1}$  & $3.79\times10^{-4}$ & $1.23\times10^{-2}$ & $3.061\%$ \\ \hline	
			23.\ \cite{ashraf2013improved} & $5.30\times10^{-2}$  & $6.00\times10^{-2}$  & $6.47\times 10^{1}$ & $9.81$  & $2.09\times10^{-2}$ & $6.42\times10^{-1}$ & $3.252\%$ \\ \hline
			24.\ \cite{kuang2019parametrically} & $1.38 \times 10^{-2}$ & $3.82 \times 10^{-2}$  &   $1.13\times 10^{2}$  & $4.91$ &    $3.60\times 10^{-3}$        &  $9.29 \times 10^{-2}$ &    $3.875\%$ \\ \hline
			25.\ \cite{nammari2018fabrication} & $1.30 \times 10^{-2}$ & $8.00 \times 10^{-2}$  &   $9.74\times 10^{1}$  & $9.81$ &    $1.34\times 10^{-2}$        &  $3.16 \times 10^{-1}$ &    $4.227\%$ \\ \hline
			26.\ \cite{dechant2017low} & $1.64 \times 10^{-2}$ & $6.00 \times 10^{-3}$  &   $1.07\times 10^{2}$  & $9.81$ &    $1.55\times 10^{-3}$        &  $3.28 \times 10^{-2}$ &    $4.727\%$ \\ \hline
			27.\ \cite{halim2014theoretical} & $4.36 \times 10^{-3}$ & $7.00 \times 10^{-3}$  &   $9.11\times 10^{1}$  & $6.00$ &    $3.78\times 10^{-4}$        &  $5.31 \times 10^{-3}$ &    $7.118\%$ \\ \hline
			28.\ \cite{gu2011impact} & $8.00\times10^{-3}$  & $1.00\times10^{-2}$ & $5.09\times 10^{1}$ & $3.90$  & $4.30\times10^{-4}$ & $5.12\times10^{-3}$ & $8.398\%$ \\ \hline
			29.\ \cite{munaz2013study} & $1.15\times10^{-2}$  & $8.00\times10^{-2}$  & $3.77\times 10^{1}$ & $4.91$  & $4.84\times10^{-3}$ & $5.42\times10^{-2}$ & $8.938\%$ \\ \hline			
			30.\ \cite{geisler2017human} & $5.74\times10^{-3}$  & $5.00\times10^{-2}$  & $3.64\times 10^{1}$ & $1.96\times 10^{1}$  & $6.57\times10^{-3}$ & $6.53\times10^{-2}$ & $10.058\%$ \\ \hline
			31.\ \cite{foisal2012multi} & $6.80\times10^{-3}$  & $4.20\times10^{-2}$  & $5.28\times 10^{1}$ & $4.91$  & $2.37\times10^{-3}$ & $2.35\times10^{-2}$ & $10.081\%$ \\ \hline
			32.\ \cite{saravia2017hybrid} & $4.80\times10^{-2}$  & $1.00\times10^{-1}$  & $4.39\times 10^{1}$ & $9.81$  & $8.60\times10^{-2}$ & $6.59\times10^{-1}$ & $13.045\%$ \\ \hline
			33.\ \cite{erturk2011broadband} & $1.40\times10^{-2}$  & $4.00\times10^{-2}$  & $5.01\times 10^{1}$ & $4.91$  & $8.40\times10^{-3}$ & $4.40\times10^{-2}$ & $19.091\%$ \\ \hline
                34.\ \cite{moss2010low} & $3.00\times10^{-2}$  & $7.84\times10^{-4}$ & $7.10\times 10^{2}$ & $4.40$  & $5.30\times10^{-3}$ & $2.33\times10^{-2}$ & $22.747\%$ \\ \hline
			35.\ \cite{moss2011broadband} & $5.90\times10^{-2}$  & $1.80\times10^{-3}$  & $2.57\times 10^{2}$ & $4.40$  & $1.00\times10^{-2}$ & $3.82\times10^{-2}$ & $26.178\%$ \\ \hline
			36.\ \cite{haroun2015study} & $3.60\times10^{-2}$  & $1.00\times10^{-1}$  & $1.38\times 10^{2}$ & $1.91\times 10^{2}$  & $8.80$ & $3.02\times10^{1}$ & $29.139\%$ \\ \hline
			\end{tabular}
		\label{table}
		
	\end{table}
	
\section{Summary and conclusions}
\label{Summary}

In this paper we have presented a theoretical upper bound on the efficacy of inertially driven energy harvesters subjected to unidirectional sinusoidal base excitation. The energy harvester is assumed to be confined within a casing, and to have no interactions with external materials or force fields. The construction of the device inside the casing can be quite arbitrary, with one or many degrees of freedom; with rigid or flexible components connected arbitrarily; with unidirectional or planar or spatial motions. Our only restrictions are that the total internal mass moving relative to the casing is $m$; the range of all internal motions in the direction of base excitation displacement is $L$; the sinusoidal base displacement has frequency $\omega$ and amplitude $A$; and all action-reaction force pairs that are relevant to the energy harvesting are internal to the casing. Usually, $A \ll L$.

For such devices, regardless of the design and construction of the device, the maximum average power extractable from the device is bounded by the maximum mechanical power input possible from the base to the casing; and that power is bounded above by
$\displaystyle \frac{mLA\omega^3}{\pi}$.

Previous authors had obtained the same bound for a single degree of freedom unidirectional oscillator. Our first contribution lies in showing theoretically that the same bound holds for a wide range of multi-degree-of-freedom devices.
A smaller theoretical contribution of our paper lies in demonstrating that for a whirling pendulum design with a simple generator model, the upper bound is $\pi/4$ times smaller than the bound for more general devices. The same reduction is known for unidirectional linear resonant oscillators. Further, for such resonant oscillators, if the inherent mechanical damping is high, then the power extracted becomes quadratic instead of linear in amplitude $A$, and efficiency of energy extraction decreases rapidly.
Finally, we have examined 36 experimental realizations of energy harvesters reported in the literature and found that the power extracted ranges from 0.0036\% to 29\% of the theoretical upper bound. Of these 36 studies, 20 reported efficiencies below 2\%, and only 3 reported efficiencies above 20\%. Based on these performance data, we suggest as tentative guidelines that an average power extraction in excess of
$\displaystyle \frac{0.2 mLA\omega^3}{\pi}$ may be considered excellent, while power extraction in excess of
$\displaystyle \frac{0.3 mLA\omega^3}{\pi}$ may be considered challenging.

We suggest that future authors reporting on experimental realizations of other energy harvesters may wish to compare their observed power output with the theoretical upper bound presented in this paper, as well as with the tentative guidelines proposed above.
	
	\section{Acknowledgments}
	AC thanks Marcelo Savi, Atanu Mohanty and Sumit Basu for discussions. CPV thanks Thomas Uchida and Ajinkya Desai for
helpful comments on earlier drafts.

\bibliography{ReferencesEHCPV}

\end{document}